\documentclass[letterpaper]{article} 
\usepackage{aaai24}  
\usepackage{times}  
\usepackage{helvet}  
\usepackage{courier}  
\usepackage[hyphens]{url}  
\usepackage{graphicx} 
\usepackage{natbib}  
\usepackage{caption} 
\usepackage{algorithm}
\usepackage{algorithmic}

\usepackage{newfloat}
\usepackage{listings}
\DeclareCaptionStyle{ruled}{labelfont=normalfont,labelsep=colon,strut=off} 
\floatstyle{ruled}
\newfloat{listing}{tb}{lst}{}
\floatname{listing}{Listing}
\usepackage{amsmath}
\usepackage{amssymb}
\usepackage{multirow}
\usepackage{xcolor}

\title{Point Transformer with Federated Learning for Predicting Breast Cancer HER2 Status from  Hematoxylin and Eosin-Stained Whole Slide Images}

\author{
    Bao Li\textsuperscript{\rm 1, 2}\equalcontrib,
    Zhenyu Liu\textsuperscript{\rm 2}\equalcontrib,
    Lizhi Shao\textsuperscript{\rm 2},
    Bensheng Qiu\textsuperscript{\rm 1},
    Hong Bu\textsuperscript{\rm 3}\thanks{Corresponding author.} ,
    Jie Tian\textsuperscript{\rm 1, 2, 4}\footnotemark[2]\\
}
\affiliations{

    \textsuperscript{\rm 1}{Center for Biomedical Imaging, University of Science and Technology of China, Hefei, China}\\
    \textsuperscript{\rm 2}{CAS Key Laboratory of Molecular Imaging, Beijing Key Laboratory of Molecular Imaging,\\
    Institute of Automation, Chinese Academy of Sciences, Beijing, China}\\
   \textsuperscript{\rm 3}{Department of Pathology, West China Hospital, Sichuan University, Chengdu, China}\\
    \textsuperscript{\rm 4}{Key Laboratory of Big Data-Based Precision Medicine, Ministry of Industry and Information Technology,\\ School of Engineering Medicine, Beihang University, Beijing, China}\\
    libao1506@mail.ustc.edu.cn, bqiu@ustc.edu.cn, hongbu@scu.edu.cn, \{zhenyu.liu, lizhi.shao, jie.tian\}@ia.ac.cn

}

\begin{document}

\maketitle

\begin{abstract}
Directly predicting human epidermal growth factor receptor 2 (HER2) status from widely available hematoxylin and eosin (HE)-stained whole slide images (WSIs) can reduce technical costs and expedite treatment selection. Accurately predicting HER2 requires large collections of multi-site WSIs. Federated learning enables collaborative training of these WSIs without gigabyte-size WSIs transportation and data privacy concerns. However, federated learning encounters challenges in addressing label imbalance in multi-site WSIs from the real world. Moreover, existing WSI classification methods cannot simultaneously exploit local context information and long-range dependencies in the site-end feature representation of federated learning. To address these issues, we present a point transformer with federated learning for multi-site HER2 status prediction from HE-stained WSIs. Our approach incorporates two novel designs. We propose a dynamic label distribution strategy and an auxiliary classifier, which helps to establish a well-initialized model and mitigate label distribution variations across sites. Additionally, we propose a farthest cosine sampling based on cosine distance. It can sample the most distinctive features and capture the long-range dependencies. Extensive experiments and analysis show that our method achieves state-of-the-art performance at four sites with a total of 2687 WSIs. Furthermore, we demonstrate that our model can generalize to two unseen sites with 229 WSIs. Code is available at: \mbox{https://github.com/boyden/PointTransformerFL}

\end{abstract}

\begin{figure}[ht]
\centering
\includegraphics{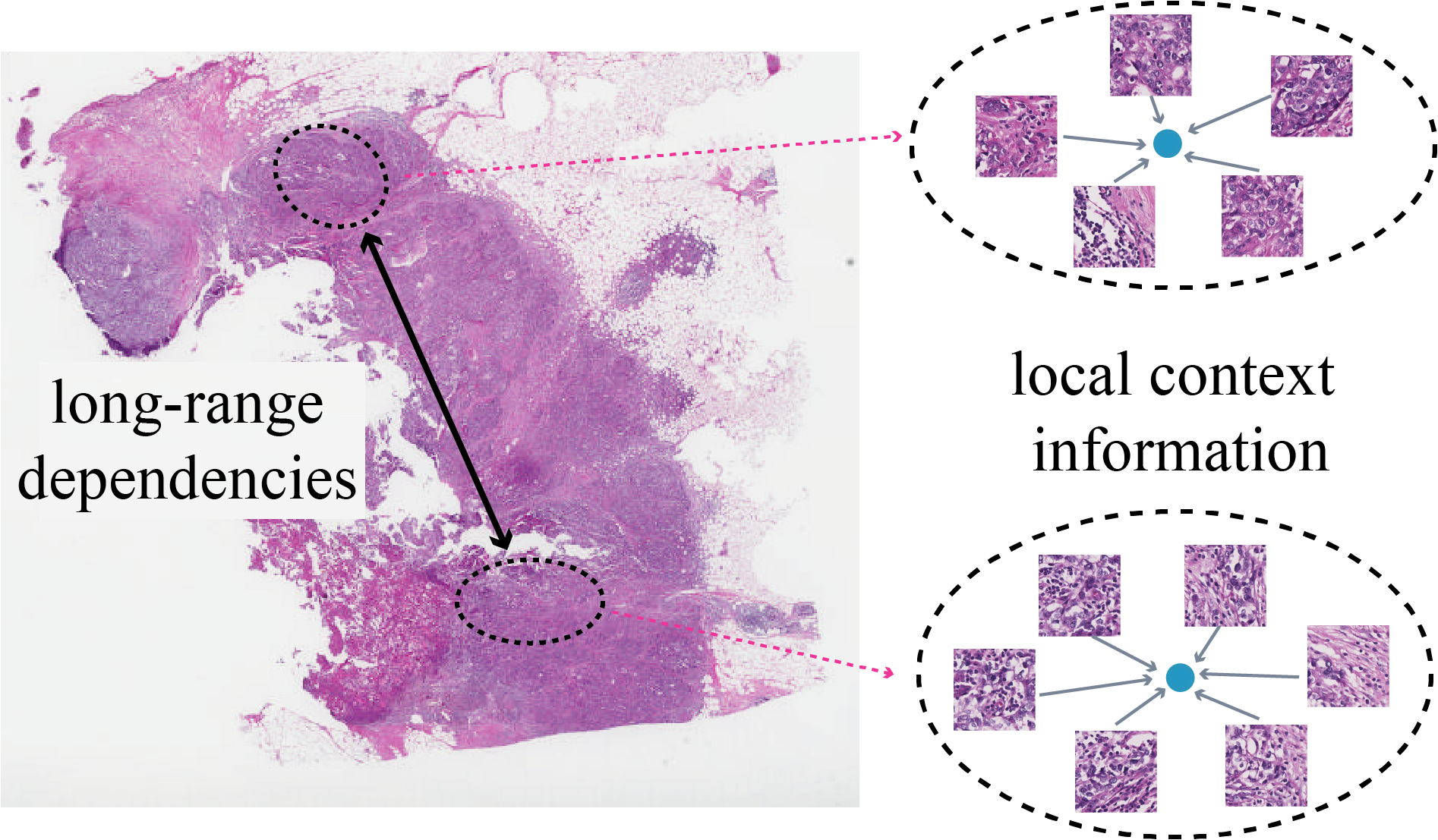} 
\caption{Local context information and long-range dependencies are 
 both essential for WSI analysis}
\label{fig:wsi_pro}
\end{figure}

\section*{Introduction}
Hematoxylin and eosin (HE)-stained whole slide images (WSIs) are now being used beyond visible tasks by applying deep learning methods~\cite{Lu2021clam, shap2021transmil, li2022deep}. These images contain subtle molecular characteristics that can be inferred using deep learning~\cite{kather2020pan, farahmand2022deep, lu2022slidegraph}. In breast cancer, accurately predicting human epidermal growth factor receptor 2 (HER2) status is crucial for guiding anti-HER2 treatment decisions ~\cite{Oh2019HER2targetedT}. Routinely, pathologists rely on HE-stained WSIs for breast cancer diagnosis, followed by specialized immunohistochemistry (IHC) and/or costly in-situ hybridization (ISH) techniques~\cite{her2test} to determine HER2 status. By utilizing deep learning, we can predict HER2 status from broadly accessible HE-stained WSIs without requiring IHC and/or ISH.

Achieving better WSI-level prediction requires large amounts of WSIs. Federated learning (FL)~\cite{mcmahan2017communication} has already exhibited promising progress in WSI analysis~\cite{lu2022histfl, jiang2022harmofl, ogier2023federated}. It can incorporate a large amount of multi-site WSIs without actual transportation of gigabyte-size WSIs and reduce the risk of data leakage. However, real-world WSIs exist non-independent and identically distributed (non-i.i.d.) scenarios. For HER2 classification, labeling imbalance and varying histological specimen preparation at different sites can adversely affect the overall performance. Although many studies have addressed the non-i.i.d. challenges~\cite{Hu2020FederatedLM, guan2023federated, zhuang2023fed} in natural scenes, these methods remain a major gap in real-world WSIs compared to centralized learning.

In the site-end feature representation, WSIs are cut into patches, and these patches' local context information and long-range dependencies (as shown in Figure~\ref{fig:wsi_pro}) are essential for WSI-level prediction, such as HER2 prediction ~\cite{kather2020pan, lu2022slidegraph} and survival analysis~\cite{chen2021multimodal, shen2022explainable, shao2023hvtsurv}. For HER2 prediction, HER2-positive patches may cluster in many separate regions of WSIs. Existing deep learning methods either treat these patches as instances using multi-instance learning (MIL) methods ~\cite{Lu2021clam, shap2021transmil, shao2023hvtsurv}, or structure the patches into a graph using graph neural networks (GNNs)~\cite{chen2021gcn, lu2022slidegraph, hou2022h}. However, the MIL-based methods lack the ability to model the local contextual information while the graph-based methods may struggle to capture long-distance dependence~\cite{xu2018representation} and need extra edge representation. Alternatively, the point neural network~\cite{qi2017pointnet, qi2017pointnet++} can treat each patch as a point with inherent position information in the Euclidean space, and hence effectively model the local context by considering the position information. Moreover, it is permutation invariant and has demonstrated proficiency in aggregating features and representing long-range dependencies~\cite{guo2021pct, lu2022transformers}, making it well-suited for WSI analysis.

\begin{figure*}[!ht]
\centering
\includegraphics[width=0.9\textwidth]{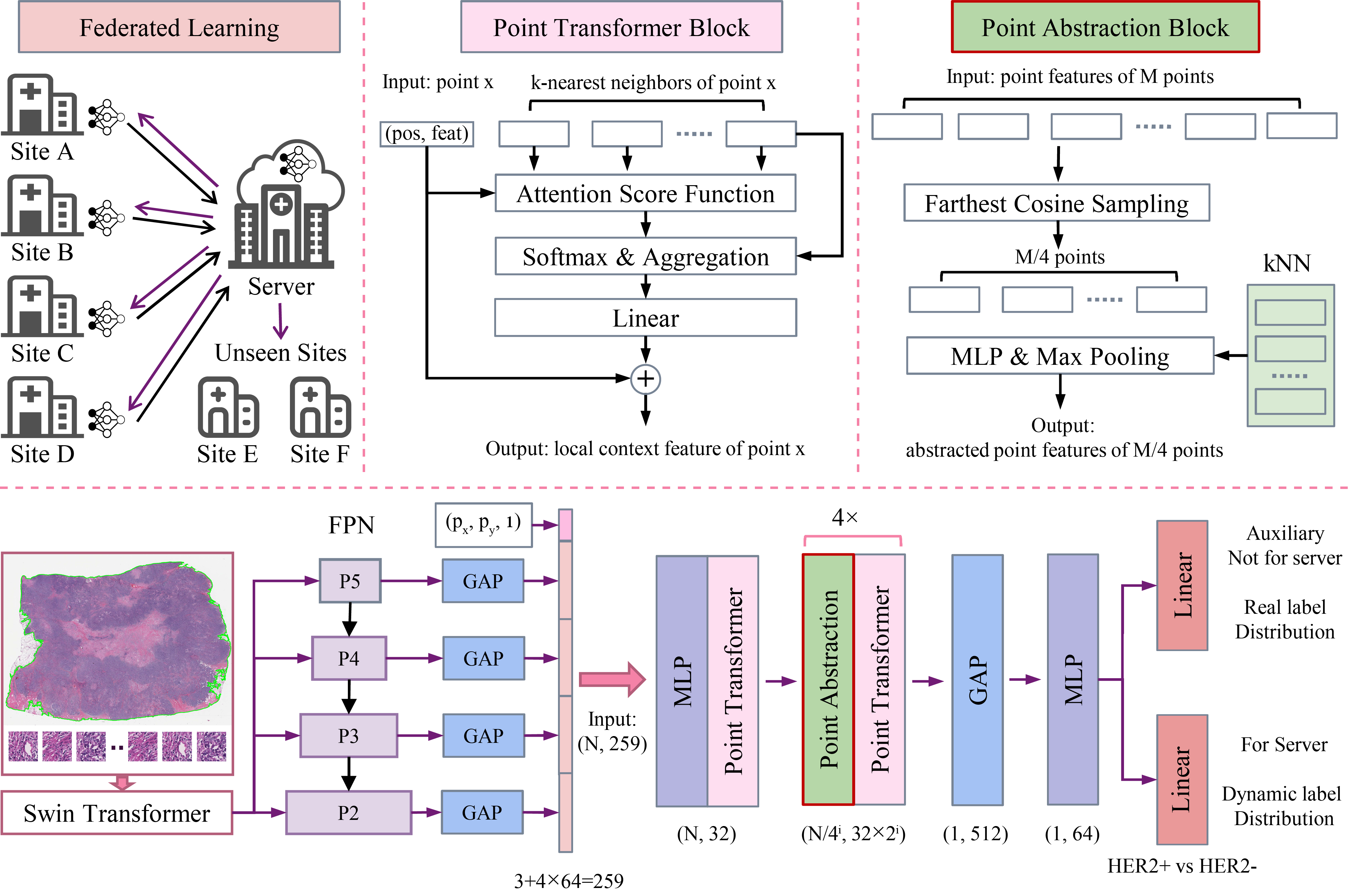}
\caption{Overview of the point transformer for predicting HER2 status from whole slide images in a federated learning framework. 4$\times$ represents that the corresponding blocks are repeated 4 times. In the $i_{th}$ block, the output shape of point features is $(N/4^i, 32\times2^i)$ and $N$ represents the total point numbers and is set to 1024. FPN: feature pyramid network, GAP: global average pooling, MLP: multilayer perceptron.}
\label{fig:model}
\end{figure*}

In this paper, we introduce a PointTransformerDDA+ to represent both the local context and the long-range dependencies. Through the point transformer block, it can capture and aggregate the local information by employing attention mechanisms enriched with position information. We also propose a novel Farthest Cosine Sampling (FCS) to capture the long-range dependencies and gather the most distinctive features based on their cosine distance. To mitigate federated learning's label-imbalance of multi-site, we present a dynamic distribution adjustment (DDA) method for a well-initialized model. It includes a distribution adjustment strategy and an auxiliary classifier. The DDA allows the resampling of labels to the same imbalance ratio initially and then dynamically adjust to the real imbalance ratio for each site without degrading the feature representation.

Our main contributions can be summarized as follows:
\begin{itemize}
    \item Unlike MIL models or graph models, we pioneer the use of point transformer for WSI analysis, which effectively captures both local context and long-range dependencies.
    \item Our proposed FCS can capture the long-range dependencies, leading to the most distinct feature aggregation.
    \item The proposed DDA mitigates the multi-site class imbalance issue, thereby enhancing model generalization.
    \item Extensive experiments on the largest WSI dataset to date for HER2 prediction in breast cancer demonstrate that our method achieves state-of-the-art performance in four sites (2687 WSIs) and two unseen sites (229 WSIs).
\end{itemize}

\section*{Related Work}
In this section, we briefly review relevant works on WSI classification and federated learning in WSI analysis.
\subsection{Whole Slide Image Classification}
Recent works have used either MIL-based or graph-based methods for WSI classification, including molecular biomarkers such as HER2 status prediction~\cite{farahmand2022deep, lu2022slidegraph}. The MIL-based methods commonly leverage attention mechanisms~\cite{ilse2018attention, chen2020pathomic, Lu2021clam} or transformers~\cite{shap2021transmil, shen2022explainable} to capture the long-distance dependence among instances. Regarding the local spatial relationship,  DSMIL~\cite{Li2020dsmil} simply extracts feature from different scales and concatenate them among scales, which do not consider the local information in a specific scale. TransMIL~\cite{shap2021transmil} models the spatial relationship among patches via transformers with conditional position encoding; however, the positions used are not based on the actual Euclidean space. Graph-based models~\cite{ham2017sage, xu2018how, lee2022derivation} are intrinsically designed to capture local information by a graph structure. In WSI analysis, Patch-GCN~\cite{chen2021gcn} regards patches as 2D point clouds while still employing GNN to analyze WSIs. SlideGraph+~\cite{lu2022slidegraph} also constructs a graph based on position information and uses edge convolution~\cite{wang2019dynamic} to model local neighbor features, leading to a SOTA HER2 prediction performance. However, the long-term dependency may limit the further improvement of GNNs in WSI analysis. While the permutation-invariant point neural network~\cite{qi2017pointnet++, zhao2021point, lu2022transformers} can capture both the local context and long-range dependencies, few studies have focused on WSI classification.
\subsection{Federated Learning in WSI Analysis}
Federated learning~\cite{mcmahan2017communication, guan2023federated} can facilitate the training of data-driven models using multi-site WSIs. HistFL~\cite{lu2022histfl} collaborates multi-sites WSI with attention MIL model and differential privacy for cancer subtype and survival prediction. Also, TNBC-FL~\cite{ogier2023federated} employs federated learning for predicting treatment outcomes in the rare subtype of breast cancer. However, general non-i.i.d. issues like skewed label distribution impedes the performance of federated learning. In natural scenes, several works replace batch normalization with group normalization~\cite{hsieh2020non}, layer normalization~\cite{du2022rethinking} or even remove the normalization layer~\cite{zhuang2023fed} to address the problems by reducing the external covariate shift. FedProx~\cite{li2020federated} introduces a regularization function to guarantee robust convergence of model in non-i.i.d. data. Additionally, FedMGDA~\cite{Hu2020FederatedLM} regards multi-site federated learning as a multi-objective optimization problem, aiming to converge to Pareto stationary solutions. With gradient normalization named FedMGDA+~\cite{Hu2020FederatedLM}, it can increase the model's robustness. Despite  the progress made in federated learning, there still remains a gap between federated learning and centralized learning in real WSIs and further improvements are still needed.

\section*{Methodology}
In this section, we start by introducing the problem definition of HER2 status prediction using a point transformer with federated learning. Then we describe the main components of the framework, including point feature extraction, point transformer block, point abstraction block, and federated learning with dynamic distribution adjustment. Figure~\ref{fig:model} illustrates the overall pipeline of our proposed framework.
\subsection{Preliminaries}
Suppose that we have $M$ sites for HER2 status prediction, our goal is to accurately determine  whether a given Whole Slide Image (WSI) is HER2-positive (HER2+) or HER2-negative (HER2-). For the $i_{th}$ site, it contains a labeled point dataset $\mathcal{P}_i=\{(\mathcal{X}_n, y_n) \mid n\in (1, ..., |\mathcal{P}_i|)\}$, where $y_n\in \{0, 1\}$ is the corresponding HER2- and HER2+ status. Within this dataset, $\mathcal{X}_n = \{x_{n, 1}, x_{n, 2}, ..., x_{n, |\mathcal{X}_n|}\}$ represents a point set in the $n_{th}$ whole slide images, where a point $x_{n,k} \in \mathbb{R}^{3+d}$ is a feature vector with 3-dim coordinates and $d$-dim point features. To determine HER2+ status from a point set $\mathcal{X}$ while considering data privacy, we employ federated learning to minimize the global cost over all sites:
\begin{align}
\mathop{\arg\min}_{\mathbf{W}} \mathcal{L}(\mathbf{W}) &= \sum_{i=1}^M \frac{|\mathcal{P}_i|}{|\mathcal{P}|} \mathcal{L}_i(\mathcal{P}_i; \mathbf{W}), 
\end{align}
where $|\mathcal{P}| = \sum_{i=1}^M|\mathcal{P}_i|$ is the total number of WSIs across all sites. For the $i_{th}$ site, the cost can be calculated by: 
\begin{align}
    \mathcal{L}_i(\mathcal{P}_i; \mathbf{W}) &= \frac{1}{|\mathcal{P}_i|}\sum_{(\mathcal{X}_k, y_k) \in \mathcal{P}_i} \ell (f_c(f_h(\mathcal{X}_k)), y_k; \mathbf{W}),
\end{align}
where $\ell$ is the loss function, $f_h:\mathcal{X} \mapsto \mathbb{R}^d$ is a point set embedding function, and $f_c:\mathcal{R}^d \mapsto \mathbb{R}$ is the final classifier function. WSIs from real sites have a general non-independent and identically distributed (non-i.i.d. ) scenario where each site has different HER2 status distributions. For the $i_{th}$ site, we denote the class imbalance ratio as $\gamma_i=\frac{|\mathcal{P}_i^-|}{|\mathcal{P}_i^+|}$, where $|\mathcal{P}_i^-|$ and $|\mathcal{P}_i^+|$ represent to the number of HER2- and HER2+ WSIs within the $i_{th}$ site.

\subsection{Point Feature Extraction}
To extract point features from a WSI, we follow the CLAM~\cite{Lu2021clam} to preprocess and patch the WSIs with details in Appendix A. Each patch is treated as a point. The corresponding coordinates for each patch are also traced and represented as a tuple$(p_x, p_y, 1)$, where 1 represents all WSIs having the same z-coordinate. Then we input the patches into a nuclei segmentation network that is pretrained using Swin-Transformer~\cite{liu2021swin} with FPN~\cite{lin2017fpn} with four levels, represented by ${P_2, P_3, P_4, P_5}$ in Figure~\ref{fig:model}. The channel of FPN is set to 64 and the outputs from all four layers of FPN are averaged and concatenated with the point coordinates as the patch-level feature $x_n\in \mathbb{R}^{3+d}$, where 3 represents the coordinates and $d=256$ represents the point feature. Thus for a WSI with a label $y$, we can obtain a point set $\mathcal{X}={x_1, x_2, ..., x_{|\mathcal{X}|}}$, where $|\mathcal{X}|$ is the number of patches in a WSI. 1024 patches or points are randomly selected with uniform distribution to reduce the memory usage and improve computational efficiency.

\subsection{Point Transformer Block}
In our approach, we adopt the original point transformer block~\cite{zhao2021point} to capture and aggregate the local context information of each point with an effective attention mechanism. For the $i_{th}$ point with its corresponding point feature $x_i$, position $p_i$. We represent its \textit{k}-nearest neighborhood points (k=16) as a subset $\mathcal{X}(i) \subset \mathcal{X}$. Then we compute the attention score between point $x_i$ and point subset $\mathcal{X}(i)$ by the attention score function $\alpha$:
\begin{align}
\alpha(x_i, x_j) = W_q(W_ix_i-W_jx_j)+PE(p_i, p_j),
\end{align}
where $x_j \in \mathcal{X}(i)$ and $PE$ is a relative position encoding function defined as:
\begin{align}
PE(p_i, p_j) = MLP(p_i-p_j).
\end{align}
MLP in the formula represents two linear layers with a ReLU activation function. 

Afterward, we aggregate the localized feature around of point $x_i$ and obtain the aggregated feature $z_i$ with a softmax function $S$. Then the output $y_i$ is computed by applying a residual connection between $x_i$ and $z_i$:
\begin{align}
z_i = \sum_{x_j \in \mathcal{X}(i)}S(\alpha(x_i, x_j))&\cdot(W_vx_j+PE(p_i, p_j)),\\
y_i = x_i& + W_zz_i.
\end{align}

\subsection{Point Abstraction Block}
To effectively reduce the cardinality of a point set and capture the long-range dependencies without missing important points, we propose a novel sampling strategy named farthest cosine sampling (FCS), as an alternative to the farthest point sampling (FPS) ~\cite{qi2017pointnet++}. 

\begin{figure}[t]
\centering
\includegraphics[width=0.4\textwidth]{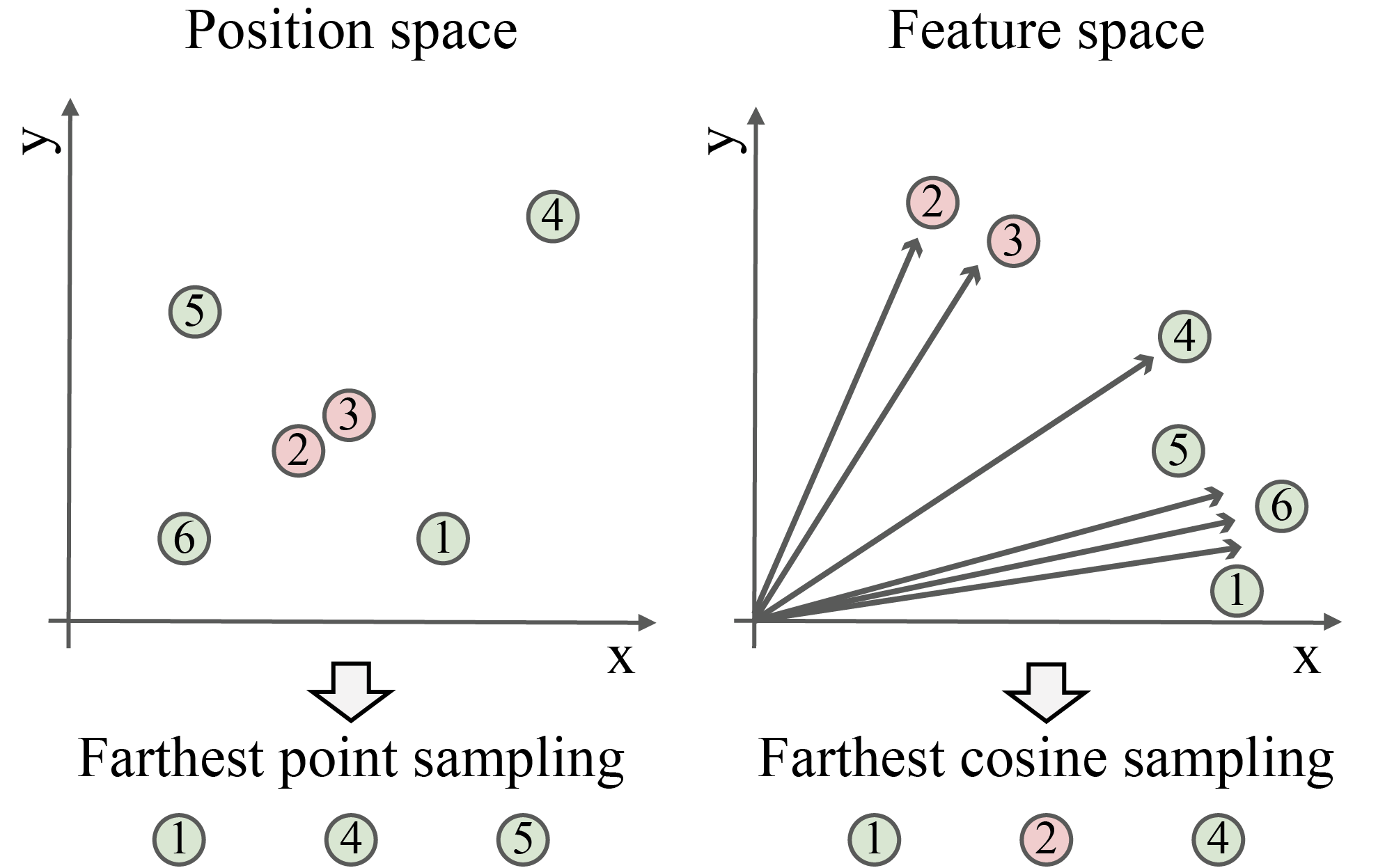} 
\caption{Difference between farthest point sampling and farthest cosine sampling. Light red: HER2+ points, light green: HER2- points.}
\label{fig:fcs}
\end{figure}

In scenarios where a WSI exhibits a majority of negative patches with only a few positive patches clustered together in a specific region, FPS may miss these positive patches, as depicted in Figure~\ref{fig:fcs}. Consequently, it can lead to false negative predictions of HER2 status. Contrary to FPS, we perform sampling in the feature space, not in the position space. For a point set $\mathcal{X}_1=\{x_1, x_2, ..., x_M\}$ with M points, we define the cosine distance as the distance metric between two points $x_i, x_j\in \mathcal{X}_1$:
\begin{align}
Dist(x_i, x_j) = 1-\frac{x_i\cdot x_j}{\max(||x_i||_2\cdot||x_j||_2, 1e-8)}.
\end{align}

\begin{algorithm}[htb]
\caption{Farthest cosine sampling}
\label{alg:fcs}
\textbf{Input}: M points with feature $\mathcal{X}_1=\{x_1, x_2, ..., x_M\}$.\\
\textbf{Output}: Sampled M/4 points $\mathcal{X}_2$.
\begin{algorithmic}[1] 
\STATE initialize an empty sampling point set $\mathcal{X}_2=\{\}$;
\STATE $x_{s,1}$ = RandomChoiceOne($\mathcal{X}_1$);
\STATE $\mathcal{X}_1 \gets \mathcal{X}_1 \setminus \{x_{s,1}\}; \mathcal{X}_2 \gets \{x_{s,1}\}$;
\WHILE{$|\mathcal{X}_2| < M/4$}
\STATE // Using cosine similarity for distance metric
\STATE $Dist(i, j)$ = $1-\frac{x_i\cdot x_{s,j}}{\max(||x_i||_2\cdot||x_{s,j}||_2, 1e-8)}$;
\STATE $\mathcal{D} = \{Dist(i, j); x_i \in \mathcal{X}_1, x_{s,j} \in \mathcal{X}_2\}$;
\STATE $x_{s} \gets \mathop{\arg\max}_{x_i\in \mathcal{X}_1} (\mathcal{D})$;
\STATE $\mathcal{X}_1 \gets \mathcal{X}_1 \setminus \{x_{s}\}; \mathcal{X}_2 \gets \mathcal{X}_2 \cup \{x_{s}\}$;
\ENDWHILE
\STATE \textbf{return} $\mathcal{X}_2$
\end{algorithmic}
\end{algorithm}

Using this distance metric, we iteratively select the $M/4$ farthest points based on Algorithm~\ref{alg:fcs}. Consequently, the FCS can effectively cover the most requisite patches 
 and capture the long-range dependencies in the feature space for better prediction of HER2 status.

After FCS, we obtain a sampled point subset $\mathcal{X}_2=\{x_{s,1}, x_{s, 2}, ..., x_{s, M/4}\}$. For each sampled point $x_i\in \mathcal{X}_2$, we define its \textit{k}-nearest neighborhood points (k=16) on $\mathcal{X}_1$ as a point subset: $\mathcal{X}_2(i) \subset \mathcal{X}_1$. Subsequently, we  group the feature from $\mathcal{X}_1$ onto $\mathcal{X}_2$ as $y_i$ for each point $x_i \in \mathcal{X}_2$ using the following equation:
\begin{align}
y_i = MaxPooling_{x_j\in \mathcal{X}_2(i)}(MLP(x_j)).
\end{align}
The MLP has two layers with each layer containing a linear transformation, batch normalization, and a ReLU activate function.

\subsection{Point Classifier Block}
After performing $4\times$ attention and abstraction operations, we obtain 4 abstract points with a grouped feature representation denoted as $F_g \in \mathbb{R}^{4\times512}$. By averaging the grouped feature, we derive the final WSI-level feature represented as $F_h\in \mathbb{R}^{64}$:
\begin{align}
    F_h = MLP(GAP(F_g)) = f_h(\mathcal{X}),
\end{align} 
where MLP has two layers with each layer containing a linear transformation and a ReLU activation function. Following the MLP, a linear layer with a softmax function named $f_c$ outputs the final HER2 status probabilities $p$. The loss is calculated using cross entropy (CE) loss function, which is formulated as:
\begin{align}
    \ell(p, y) = -\frac{1}{|\mathcal{P}|} \sum_{i}^{|\mathcal{P}|} \sum_{c=0}^1y_{c}^i\log (p_c^i).
\end{align} 

\begin{figure}[t]
\centering
\includegraphics[width=0.4\textwidth]{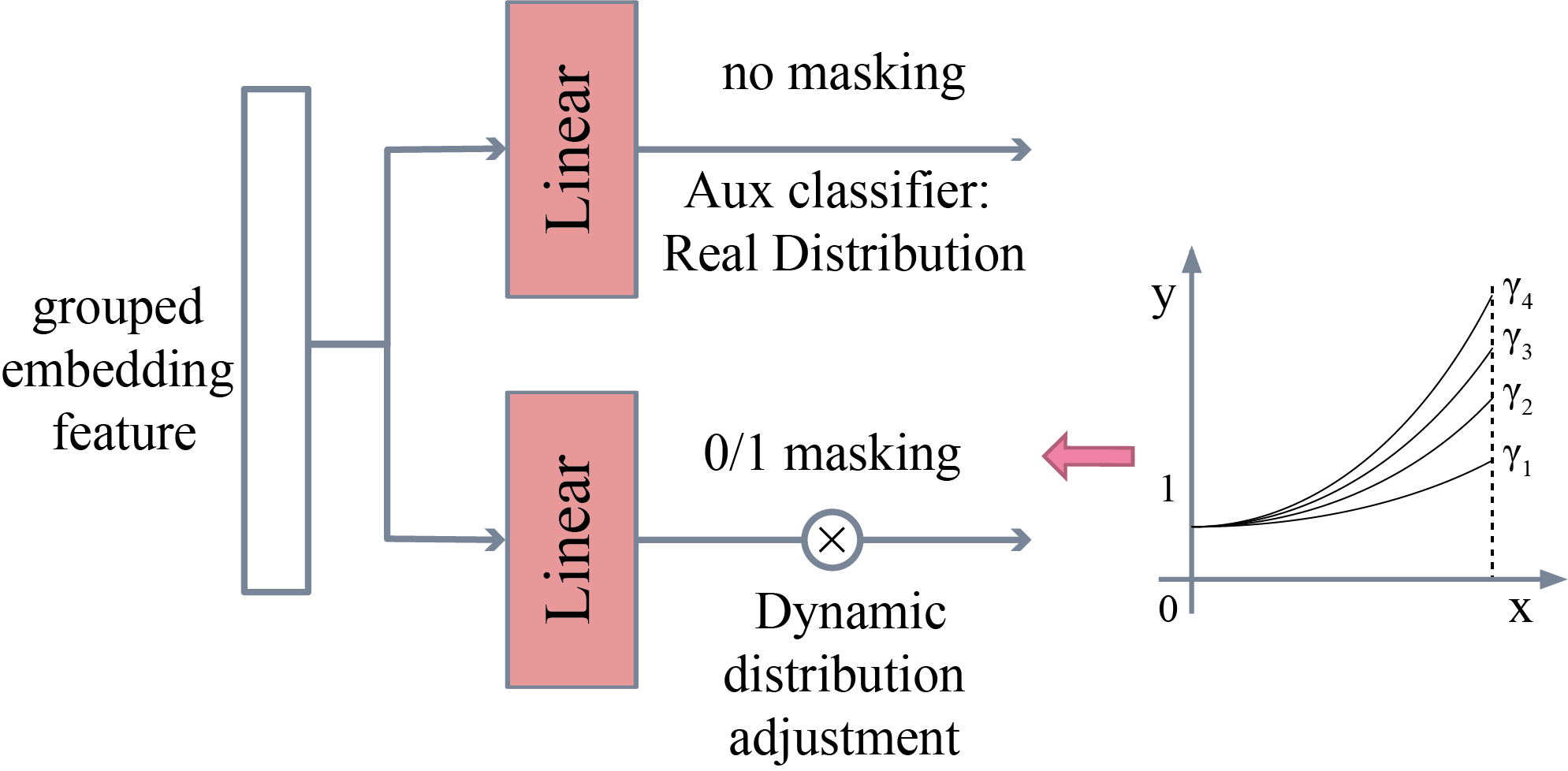} 
\caption{Dynamic distribution adjustment and auxiliary classifier for federated learning.}
\label{fig:aux}
\end{figure}

\subsection{Federated Learning with Dynamic Distribution Adjustment}
The WSIs from real sites exhibit a general non-independent and identically distributed (non-i.i.d.) scenario in which each site has different label distribution, which can impact the performance of federated learning~\cite{guan2023federated}. To mitigate this issue, we introduce a novel dynamic distribution adjustment strategy. It subsampling the majority of HER2- WSIs to the same label distribution in the initial training stage of federated learning. Then the balanced label distribution dynamically adjusts to the real distribution after progressive training. Specifically, we keep all the HER2+ WSIs and generated a 0/1 mask $M(k)$ for the HER2- WSIs. This mask is created using a Bernoulli distribution $\mathcal{B}$ with a probability of $b_k$ at $k_{th}$ iteration.
\begin{align}
    M_i(k) &= \mathcal{B}(b_i^k), 
\end{align}
\begin{align}
    b_i^k &= \frac{1}{\gamma_i} + (1-\frac{1}{\gamma_i})\cdot\frac{e^{k/K}-1}{e-1}\in [\frac{1}{\gamma_i}, 1],
\end{align}
where $\gamma_i$ is the $i_{th}$ site's imbalanced ratio and $K$ is the total optimization steps.
The loss function can be formulated as:
\begin{align}
    \mathcal{L}_{cls} &= M\cdot \ell(f_c(F_h), y)\\
    &= -\frac{1}{|\mathcal{P}_i|}[\sum_{i}^{|\mathcal{P}_i^-|} M_i(k)\log (p_0^i) + \sum_{i}^{|\mathcal{P}_i^+|} \log (p_1^i)],
\end{align}
where $|P_i^-|, |P_i^+|$ represent the number of HER2- and HER2+ WSIs.

In every iteration $k$, we note $b_i^k\gamma_i$ as the imbalance ratio involved in loss computation. Figure~\ref{fig:aux} shows that at iteration $k=0$, all sites have $b_i^k\gamma_i=1$ indicating that they share the same label distribution with an equal ratio between HER2- and HER2+. As the training progresses, the distribution gradually shifts towards each site's real distribution $\gamma_i$. By this strategy, the models are initially trained within a similar distribution, leading to a well-initialized model for accurately predicting HER2 status.

The above subsampling may arise a potential information loss problem, leading to insufficient feature representation. To address this concern, we add an auxiliary classifier that incorporates each site's real label distribution:

\begin{align}
    \mathcal{L}_{aux}(p, y) = -\frac{1}{|\mathcal{P}_i|}[\sum_{i}^{|\mathcal{P}_i^-|}&\log (p_0^i) + \sum_{i}^{|\mathcal{P}_i^+|} \log (p_1^i)],
\end{align}
\begin{align}
    \mathcal{L}_{total} &= \mathcal{L}_{cls} + \mathcal{L}_{aux}.
\end{align}
This auxiliary classifier does not involved in the weights synchronization with the server model during the training of federated learning. Instead, it serves to guarantee the quality of feature representation as previous studies have demonstrated that models trained on imbalanced data can still learn high-quality feature representations~\cite{kang2020decoupling, lee2021abc}.
Detailed pseudo codes are shown in algorithm~\ref{alg:fed}.

\section*{Experiments}
\subsection{Dataset and Experimental Settings}
We evaluate our point transformer model for breast cancer HER2 status prediction using the largest WSI from six sites with a total of 2,916 WSIs. The sites are denoted as follows: Site A (TCGA-BRCA)~\cite{cancer2012brca}, Sites B and C (internal hospitals with ethics committee approval), Site D~\cite{sousa2022hero}, and Sites E and F~\cite{farahmand2022deep, Qaiser2017HER2CC}. Four sites participate in federated learning, while the remaining two sites serve as unseen data for external tests. Sites A, B, C, and D are split into training, validation, and test sets with a ratio of 6:1:3, as shown in Table~\ref{tab: wsi_site}. The splits are repeated five times, and the best model is selected based on the validation set in each split. The mean area under the ROC curve (AUC) is reported for the test set.

We refer to the point transformer with FCS as PointTransformer+, the variant with DDA as PointTransformerDDA, and the combined variant as PointTransformerDDA+.

\begin{table}[htb]\small
\centering
\begin{tabular}{c@{\hspace{1.5mm}}|c@{\hspace{1.5mm}} c@{\hspace{1.5mm}} c@{\hspace{1.5mm}} c@{\hspace{1.5mm}} c@{\hspace{1.5mm}}|c@{\hspace{1.5mm}} c@{\hspace{1.5mm}}}
\hline
\multirow{2}{*}{WSIs} & \multicolumn{5}{c|}{Federated Sites} & \multicolumn{2}{c}{Unseen Sites}\\
& Site A & Site B & Site C & Site D & Total & Site E & Site F\\
\hline
HER2$-$  & 669 & 672 & 332 & 306 & 1979 &98 & 26 \\
HER2$+$ & 118 & 214 & 172 & 204 & 708 &93 & 12 \\
Total  & 787 & 886 & 504 & 510 & 2687 &191 & 38 \\
\hline
$\gamma$  & 5.7 & 3.1 & 1.9 & 1.5 & 2.8 & 1.1 & 2.2 \\
\hline
Train  & 472 & 532 & 302 & 306 & 1612 & - & - \\
Val    & 79  & 88  & 50  & 51  & 268  & - & - \\
Test   & 236 & 266 & 152 & 153 & 807  & 191 & 38 \\
\hline
\end{tabular}
\caption{WSIs and their HER2 status number in each site. $\gamma_i=\frac{|\mathcal{P}_i^-|}{|\mathcal{P}_i^+|}$ represents the imbalance ratio. WSIs are split into training (60\%), validation (10\%), and test (30\%) set.}
\label{tab: wsi_site}
\end{table}

\begin{table*}[!htp]\small
\centering
\begin{tabular}{lcccccc}
\hline
Experiment & Method & Average & Site A & Site B & Site C & Site D \\ \hline
\multirow{3}{*}{Ours} & PointTransformerDDA+  & \textbf{0.816} ± 0.019 & \textbf{0.766} ± 0.025 & \textbf{0.866} ± 0.021 & \textbf{0.837} ± 0.036 & \textbf{0.760} ± 0.046  \\ 
                             & PointTransformerDDA   & 0.793 ± 0.013 & 0.730 ± 0.029  & 0.855 ± 0.013 & 0.804 ± 0.024 & 0.758 ± 0.022 \\
                             & PointTransformer+     & 0.806 ± 0.015 & 0.752 ± 0.018 & 0.844 ± 0.008 & 0.823 ± 0.024 & 0.757 ± 0.043 \\ \hline
\multirow{2}{*}{Point-based} & PointTransformer [1] & 0.771 ± 0.012 & 0.717 ± 0.037 & 0.834 ± 0.026 & 0.776 ± 0.037 & 0.721 ± 0.035 \\
                             & PointNet++ [2] & 0.763 ± 0.017 & 0.696 ± 0.039 & 0.830 ± 0.033  & 0.746 ± 0.045 & 0.730 ± 0.042  \\ \hline
\multirow{4}{*}{MIL-based}   & CLAM-SB [3] & 0.767 ± 0.032 & 0.712 ± 0.044 & 0.793 ± 0.037 & 0.766 ± 0.072 & 0.748 ± 0.022 \\
                             & DSMIL [4] & 0.693 ± 0.096 & 0.647 ± 0.065 & 0.738 ± 0.103 & 0.706 ± 0.080  & 0.675 ± 0.111 \\
                             & TransMIL [5] & 0.790 ± 0.019  & 0.739 ± 0.038 & 0.824 ± 0.021 & 0.805 ± 0.036 & 0.759 ± 0.040  \\
                             & HistoFL [6] & 0.757 ± 0.039 & 0.729 ± 0.048 & 0.776 ± 0.050  & 0.759 ± 0.076 & 0.733 ± 0.012 \\ \hline
\multirow{3}{*}{Graph-based} & GraphSAGE [7] & 0.711 ± 0.026 & 0.656 ± 0.053 & 0.735 ± 0.027 & 0.692 ± 0.044 & 0.685 ± 0.021 \\
                             & Patch-GCN [8] & 0.750 ± 0.037  & 0.700 ± 0.035   & 0.768 ± 0.062 & 0.766 ± 0.030  & 0.727 ± 0.047 \\
                             & SlideGraph+ [9] & 0.783 ± 0.019 & 0.736 ± 0.029 & 0.828 ± 0.012 & 0.804 ± 0.037 & \textbf{0.785} ± 0.013\\ \hline
\end{tabular}
\caption{Comparison of our model with other point, multi-instance, and graph-based models. [1]~\cite{zhao2021point}, [2]~\cite{qi2017pointnet++}, [3]~\cite{Lu2021clam}, [4]~\cite{Li2020dsmil}, [5]~\cite{shap2021transmil}, [6]~\cite{lu2022histfl}, [7]~\cite{ham2017sage}, [8]~\cite{chen2021gcn}, [9]~\cite{lu2022slidegraph}.}
\label{table:method_res}
\end{table*}

\begin{algorithm}[htb]
\caption{Point transformer with federated learning}
\label{alg:fed}
\textbf{Input}: M participating sites with point set $\mathcal{P}_m=(\mathcal{X}_m, y_m)$\\
and label imbalanced ratio $\gamma_m$ where $m\in\{1, ..., M\}$. Optimization epochs: K, communication pace: E.\\
\textbf{Output}: Model's weights $W_s$. 
\begin{algorithmic}[1] 
\STATE initialize the point transformer's function: $f_h, f_c, f_{aux}$;
\STATE initialize the same weights for the server and sites: $W_s^0, W_{s,1}^0, ..., W_{s,M}^0$;
\FOR{$k=0$ \TO $K-1$}
    \FOR{$m=1$ \TO $M$}
        \STATE $F_h = f_h(\mathcal{X}_m)$;\\
        \STATE // Dynamic distribution adjustment
        \STATE $M_m = \mathcal{B}(\frac{1}{\gamma_m}+(1-\frac{1}{\gamma_m})\cdot\frac{e^{k/K}-1}{e-1})$;
        \STATE $\mathcal{L}_{m} = M_m\cdot\ell
        (f_c(F_h), y_m) + \ell(f_{aux}(F_h), y_m)$;\\
        \STATE // Local site update
        \STATE $W_{s,m}^{k+1} \gets W_{s,m}^{k} - lr \cdot \nabla_{W_{s,m}^{k}} \mathcal{L}_m$;
    \ENDFOR
    \IF{$(k+1) \mod  E = 0$}
        \STATE // Server site update \\
        \FOR{each layer $l$ in point transformer}
        \IF{$l$ is not the auxiliary classifier's layer}
        \STATE $W_s^{k+1, l} \gets \sum_{m=1}^M{\frac{|\mathcal{P}_i|}{|\mathcal{P}|}W_{s,m}^{k+1, l}}$;\\
        \STATE $W_{s,m}^{k+1, l} \gets W_s^{k+1, l}$;\\
        \ENDIF
        \ENDFOR
    \ENDIF
\ENDFOR
\STATE \textbf{return} $W_s$
\end{algorithmic}
\end{algorithm}

\subsection{Implementation Details}
Our models are implemented using PyTorch 1.12.0 on a workstation with an RTX 3090 GPU. The models are trained for 200 epochs with a learning rate of 1e-3 and L2 regularization of 1e-5. Point data augmentation is used with details in Appendix B. The learning rate warm-up is tuned for the first 10 epochs, followed by a cosine decay scheduler. The Adam optimizer is adopted for weight updates.

\subsection{Comparison with WSI Classification Methods}
We include PointNet++~\cite{qi2017pointnet++}, MIL-based models: CLAM-SB~\cite{Lu2021clam}, DSMIL~\cite{Li2020dsmil}, TransMIL~\cite{shap2021transmil}, HistFL\cite{lu2022histfl}, and graph-based models: GraphSAGE~\cite{ham2017sage}, Patch-GCN~\cite{chen2021gcn}, SlideGraph+~\cite{lu2022slidegraph} for comprehensive model comparison. All of the compared models are implemented with federated average settings.

Table~\ref{table:method_res} shows that point-based models offer a competitive performance compared to MIL-based and graph-based methods. The point-based models process a unique advantage by effectively integrating both the local neighborhood features, similar to graph-based methods, and capturing the long-range dependencies, similar to MIL-based models. By introducing the novel FCS or/and DDA strategy, the point transformer achieves better performance compared to other models and PointTransformerDDA+ achieve the start-of-the-art AUC in the test set and three federated sites. Of note that TransMIL~\cite{shap2021transmil} also offers a high AUC compared to other related models. TransMIL also incorporates position encoding in the model, indicating that point position contributes to improved performance in predicting HER2 status. Further analysis of position information can be found in Appendix C.

\begin{table}[t]\small
\centering
\begin{tabular}{l@{\hspace{1mm}}c@{\hspace{1mm}}c@{\hspace{1mm}}c@{\hspace{1mm}}c@{\hspace{1mm}}c@{\hspace{1mm}}}

\hline
Federated Setting & Average & Site A & Site B & Site C & Site D \\
\hline
Centralization       & 0.823 & 0.722 & 0.839 & 0.824 & 0.819 \\
PointTransformerDDA+ & \textbf{0.816} & \textbf{0.766} & \textbf{0.866} & \textbf{0.837} & \textbf{0.760}  \\
PointTransformerDDA  & 0.793 & 0.730 & 0.855 & 0.804 & 0.758 \\
PointTransformer+    & 0.806 & 0.752 & 0.844 & 0.823 & 0.757 \\
\hline
FedAVG [1]              & 0.771 & 0.717 & 0.834 & 0.776 & 0.721 \\
FedGroupNorm [2]       & 0.783 & 0.733 & 0.832 & 0.775 & \textbf{0.768} \\
FedProx [3]          & 0.788 & 0.759 & 0.836 & 0.800 & 0.719 \\
FedMGDA [4]          & 0.773 & 0.723 & 0.818 & 0.773 & 0.741 \\
FedMGDA+ [4]         & 0.780 & 0.734 & 0.813 & 0.804 & 0.738 \\
FedWon [5]        & 0.774 & 0.716 & 0.824 & 0.777 & 0.728 \\
\hline
\end{tabular}
\caption{Comparison of different federated learning settings. [1]~\cite{mcmahan2017communication}, [2]~\cite{hsieh2020non}, [3]~\cite{li2020federated}, [4]~\cite{Hu2020FederatedLM}, [5]~\cite{zhuang2023fed}.}
\label{table:fed}
\end{table}

\begin{table}[!ht]\small
\centering
\begin{tabular}{lcc}
\hline
Unseen   Sites          & Site   E & Site F \\
\hline
PointTransformerDDA+ & 0.793   & 0.791 \\
PointTransformerDDA  & 0.795   & 0.802 \\
PointTransformer+    & \textbf{0.800}   & \textbf{0.806} \\
\hline
\end{tabular}
\caption{Performance of the point transformer in the unseen sites.}
\label{table:unseen}
\end{table}

\subsection{Comparison with Federated Learning Methods}
We also represent the point transformer's performance with different federated learning methods. Table~\ref{table:fed} shows that our proposed PointTransformerDDA+ achieves the best total AUC among other methods and is the closest to the centralized training. Among the two proposed strategies FCS and DDA, FCS can capture the most discriminative features and DDA can mitigate the issue of non-i.i.d scenario; both of them can lead to a better federated learning performance. While GroupNorm~\cite{hsieh2020non} reaches higher performance at Site D, we observe that GroupNorm is sensitive to our data and relies on careful fine-grained group number selection with details in Appendix D. Moreover, although our models are not specifically designed for unseen scenarios, they still achieve commendable performance (AUC $>$ 0.79) for two unseen sites, as shown in Table~\ref{table:unseen}.

\begin{table}[!ht]\small
\centering
\begin{tabular}{l@{\hspace{3mm}}c@{\hspace{3mm}}c@{\hspace{3mm}}c@{\hspace{3mm}}c@{\hspace{3mm}}c@{\hspace{3mm}}c@{\hspace{3mm}}}
\hline
Method & FCS & Average & Site A & Site B & Site C & Site D \\
\hline
\multirow{2}{*}{DDA} & $\times$ & 0.793 & 0.730  & 0.855 & 0.804 & 0.758  \\
& \checkmark   & \textbf{0.816} & \textbf{0.766}  &\textbf{ 0.866} & \textbf{0.837} & \textbf{0.760}   \\
\hline
\multirow{2}{*}{Base} & $\times$ & 0.771 & 0.717  & 0.834 & 0.776 & 0.721  \\
& \checkmark   & \textbf{0.806} & \textbf{0.752}  & \textbf{0.844} & \textbf{0.823} & \textbf{0.757}  \\
\hline
\multirow{2}{*}{NoFed} & $\times$ & - & 0.639 & 0.799 & 0.687 & 0.728 \\
& \checkmark   & - & \textbf{0.659}  & \textbf{0.831} & \textbf{0.711} & \textbf{0.729} \\
\hline
\end{tabular}
\caption{Ablation experiments of farthest cosine sampling (FCS). NoFed: the model is trained at each site locally.}
\label{table:point_ablation}
\end{table}

\begin{table}[!ht]\small
\centering
\begin{tabular}{l@{\hspace{1mm}}cccc}
\hline
IHC Score 2+               & Average  & Site   A & Site   B & Site   D \\
\hline
PointTransformerDDA+ & 0.712 & 0.688   & 0.733   & 0.723   \\
PointTransformerDDA  & 0.703 & 0.668   & 0.710   & 0.735   \\
PointTransformer+    & 0.747 & 0.726   & 0.748   & 0.730  \\
\hline
\end{tabular}
\caption{The AUC of our model in the IHC score 2+ subset. Site C is excluded due to no IHC 2+ WSIs .}
\label{table:her2}
\end{table}

\subsection{Ablation Studies}
\subsubsection{Farthest cosine sampling} We first evaluate the effectiveness of FCS in the DDA and base federated average settings. We also assess FCS's impact at each site by training locally rather than employing a federated learning scheme. Table~\ref{table:point_ablation} shows that FCS can consistently bring improved performance in all settings, including local training without federated learning. FCS can capture more discriminative features, leading to better performance. Of note, FCS only brings little benefit for HER2 status in Site D. This could be attributed to the fact that Site D primarily consists of biopsy WSIs with a lower number of total points compared to other sites. Consequently, the use of FPS sampling is sufficient to cover almost all the patches in Site D.

\subsubsection{Percentage of training WSIs} 
\begin{figure}[t]
\centering
\includegraphics[width=0.4\textwidth]{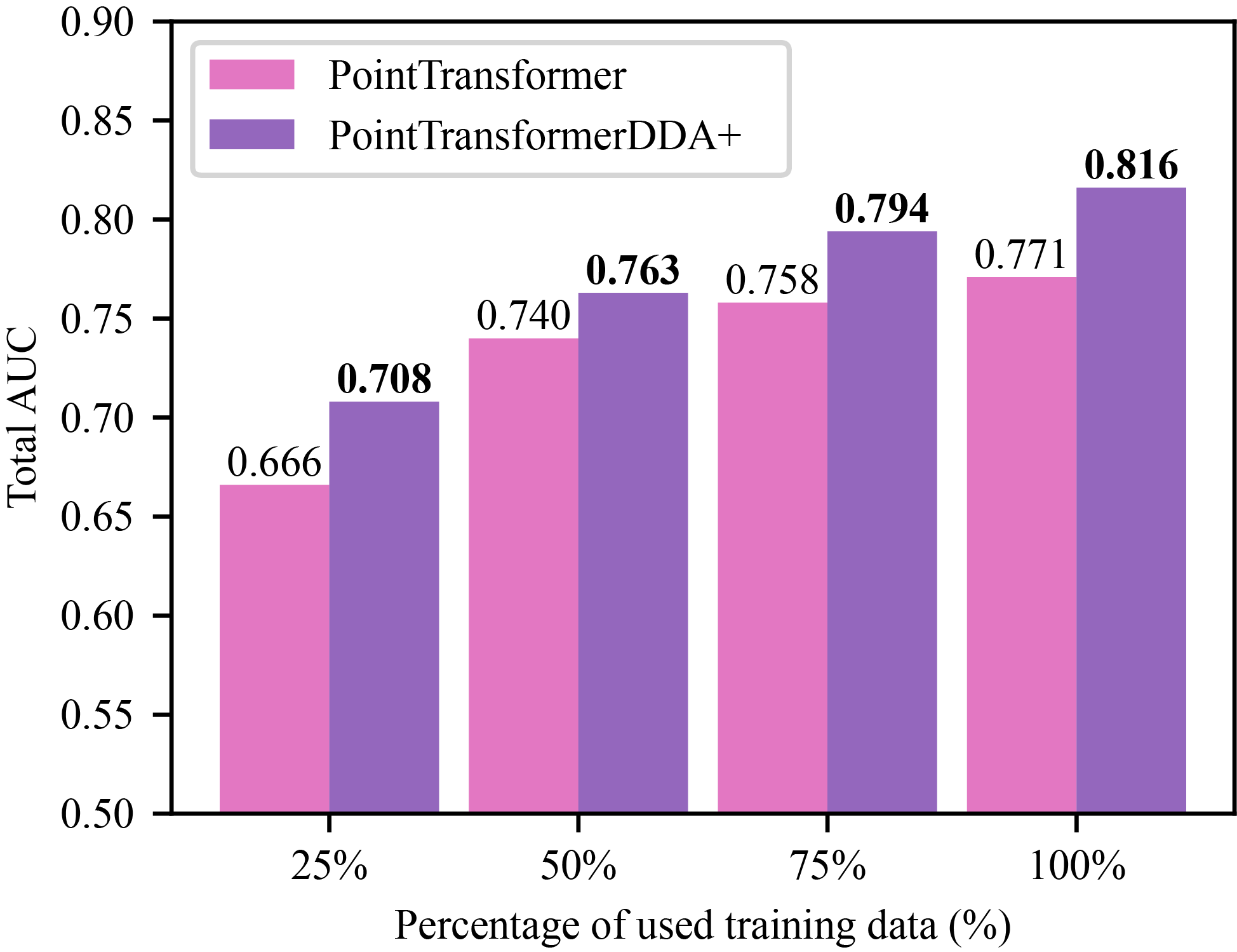} 
\caption{Performance compassion with different percentages of training WSIs.}
\label{fig:data_ratio}
\end{figure}

Our model contains the largest number of WSIs to date for predicting HER2 status. However, real-world scenarios may lack sufficient WSIs. Therefore, we evaluate our model's performance by reducing the training WSIs to 75\% (1209 WSIs), 50\% (806 WSIs), 25\% (403 WSIs). We exclude the use of 10\% of the training WSIs as it results in less than 10 positive WSIs at each site, making it unsuitable for this experiment. Figure~\ref{fig:data_ratio} shows that our model outperforms the base PointTransformer in all settings. The PointTransformerDDA+ with 50\% of training WSIs achieve an AUC that is merely 0.008 lower than the PointTransformer model (0.763 vs 0.771) trained with 100\% of the training WSIs. Moreover, with only 25\% of training WSIs, the performance of PointTransformerDDA+ still outperforms DSMIL~\cite{Li2020dsmil} (0.708 vs 0.693).

\subsubsection{IHC2+ Subset Analysis}
In real clinical scenarios IHC score 2+, pathologists cannot assess the HER2 status from IHC and require further expensive ISH tests. However, Table~\ref{table:her2} shows that our model still achieves impressive performance with an average AUC $>$ 0.7 in the test set. It can bring us an opportunity to reduce the reliance on ISH tests, thereby offering cost savings and faster HER2 status assessment.

\section*{Conclusion}
Unlike MIL-based or graph-based methods, we regard a WSI as a point cloud with position information to derive the HER2 status from HE-stained WSIs, highlighting the effectiveness of point neural networks for WSI analysis. Specifically, a farthest cosine sampling is proposed to capture the long-range dependencies and aggregate most discriminative point features. Additionally, when utilizing federated learning, we proposed a dynamic distribution adjustment to mitigate the non-i.i.d. scenario of label imbalance in real-world WSIs. Extensive experiments have demonstrated the efficacy of our two components. Our models further achieve impressive performance in both unseen sites and IHC score 2+ subsets. 

\section*{Acknowledgments}
This work was supported by the National Natural Science Foundation of China (No. 62333022) and the Beijing Natural Science Foundation (No. JQ23034).

\bibliography{aaai24}
\clearpage
\appendix

\paragraph{Roadmap of Appendix:} The Appendix includes details of WSI prepossess, WSI dataset and further experimental analysis of point transformer, and federated learning in our model.

\section{WSI Prepossess and Feature Extraction}
\label{app:wsi}
We follow CLAM~\cite{Lu2021clam} to remove the white background region and then cut the tissue regions into patches with a fixed size of 256$\times$256 at 20$\times$ magnification (~0.5$\mu m$/pixel) without overlap. For some WSIs that may not contain 20$\times$ magnification. For example, a WSI may only have 40$\times$, 10$\times$ magnification, we crop a larger size (40$\times$: 512$\times$512) or smaller size (10$\times$: 128$\times$128) and then resized to 256$\times$256 to guarantee the same field of view (FOV). The number of patches per WSI ranges from thousands to tens of thousands. The feature extraction in our study involved training of a nuclei segmentation model using the public nuclei dataset PanNuke~\cite{gamper2019pannuke, gamper2020pannuke}. The nuclei segmentation model is trained under an updated version of HTC-Lite from ~\cite{wang2021seesaw} and is implemented using an open-source MMdetection toolbox.

\section{Details of WSI Dataset}
\label{app:wsi_data}
The WSIs used in our study are involved with four publicly available datasets (Sites A, D, E, and F) and two internal datasets (Sites B and C) with the ethics committee's approval. Site A is a well-known project TCGA-BRCA~\cite{cancer2012brca}, and is widely used for WSI analysis. Site B's WSIs are retrospectively collected from West China Hospital of Sichuan University. Site C's WSIs are from out-of-hospital consultation cases of Sichuan University West China Hospital. Site D is the HEROHE Challenge~\cite{sousa2022hero} dataset, which mainly contains biopsy WSIs, and Sites E and F are the Yale HER2~\cite{farahmand2022deep} dataset and the HER2C~\cite{Qaiser2017HER2CC} datasets, respectively. All of the public datasets can be obtained from the public website offered in each paper. Note that Site F or HER2C dataset only includes an IHC score for each WSI without corresponding ISH score for IHC 2+. Thus we filter out IHC 2+ in Site F and regard the IHC 0/1+ as HER2- and IHC 3+ as HER2+. 

During the training process, various data augmentation techniques are employed to augment the point data. These include random point dropout, point jitter, scaling in [0.8, 1.25], and shifting with a range of 0.1.

\section{Further Analysis of PointTransformer}
\label{app:point_trans}
The PointTransformer involves the random selection of 1024 points and the farthest cosine sampling, which can introduce a degree of randomness into the model's results. Here we randomly repeat our model with 100 or even 1000 times to assess the stability of our model's prediction. As shown in Table~\ref{table:repeat}, we observe a 0.37\% standard deviation in the test AUC, and the difference between the maximum and minimum AUC is only 0.022. These results demonstrate the robustness of our model to random sampling and indicate its consistent and reliable performance.

\begin{table}[!ht]\small
\centering
\setlength{\tabcolsep}{3mm}
\begin{tabular}{ccccc}
\hline
Repeat & Mean & Min & Max & Max-Min \\
\hline
100    & 0.805±0.0037 & 0.796 & 0.817 & 0.021   \\
1000   & 0.806±0.0036 & 0.795 & 0.817 & 0.022  \\
\hline
\end{tabular}
\caption{Effects of randomness in PointTransformerDDA+.}
\label{table:repeat}
\end{table}

\begin{table}[!ht]\small
\centering
\setlength{\tabcolsep}{2mm}
\begin{tabular}{cccccc}
\hline
Position    & Average & Site A & Site   B & Site C & Site D \\
\hline
0           & 0.700 & 0.670  & 0.721    & 0.704  & 0.676  \\
1           & 0.709 & 0.663  & 0.733    & 0.698  & 0.714  \\
Pos         & 0.765 & 0.707  & 0.805    & \textbf{0.787}  & 0.696  \\
Pos + Aug & \textbf{0.771} & \textbf{0.717} & \textbf{0.834} & 0.776  & \textbf{0.721}  \\
\hline
\end{tabular}
\caption{The effects of position information and point data augmentation in PointTransformer.}
\label{table:point_pos}
\end{table}

Table~\ref{table:point_pos} shows that point data augmentation can increase HER2 prediction performance (AUC: 0.771 vs 0.765). Note that point augmentation is another feature that does not exist in the existing methods. We also evaluate the position effect by replacing the point position information with 0 or 1. In Table~\ref{table:point_pos}, we can observe that the model's performance significantly decreases when position information is omitted. This highlights the importance of incorporating point position information for WSI analysis.

\section{Further Analysis of Federated Learning}
\label{app:fd}
In this section, we first evaluate the effect of group normalization's number selection. Table~\ref{table:group} shows that our data is very sensitive to the choice of group number. Extreme group numbers (1: Layer Normalization, Max: Instance Normalization) and small group numbers lead to poor performance. On the other hand, group numbers 8 or 32 exhibit improved performance. Therefore, while group normalization can yield favorable results, it requires careful selection of the group number for our real-world WSIs. 

\begin{figure}[!ht]
\centering
\includegraphics[width=0.46\textwidth]{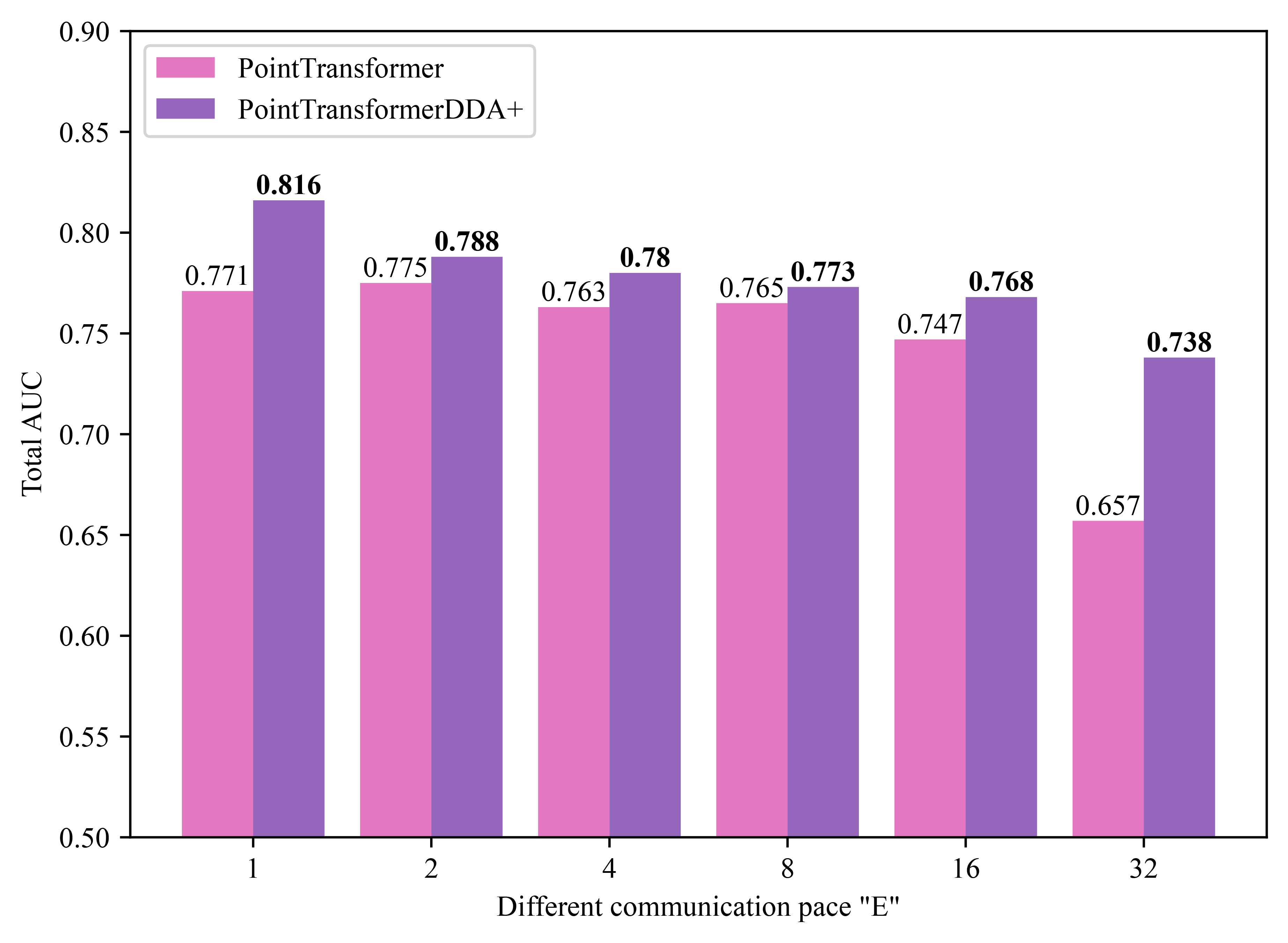} 
\caption{Performance compassion with different communication paces E.}
\label{fig:fed_E}
\end{figure}

\begin{table}[!ht]\small
\centering
\setlength{\tabcolsep}{1mm}
\begin{tabular}{cccccc}
\hline
Group   Number          & Average & Site A & Site   B & Site C & Site D  \\
\hline
1 (Layer Norm)          & 0.688 & 0.638 & 0.700 & 0.694 & 0.678 \\
2                       & 0.595 & 0.528 & 0.635 & 0.579 & 0.607 \\
4                       & 0.659 & 0.586 & 0.655 & 0.626 & 0.653 \\
8                       & \textbf{0.783} & 0.733 & 0.832 & 0.775 & 0.768 \\
16                      & 0.757 & 0.720 & 0.805 & 0.738 & 0.770 \\
32                      & 0.782 & 0.755 & 0.833 & 0.778 & 0.767 \\
64                      & 0.771 & 0.739 & 0.828 & 0.775 & 0.754 \\
MAX   (Instance Norm) & 0.532 & 0.523 & 0.538 & 0.529 & 0.542 \\
\hline
\end{tabular}
\caption{Performance of the point transformer using group normalization with different group numbers.}
\label{table:group}
\end{table}
Considering the variations in hardware and network quality, different sites may experience delays in communication frequency. We simulate a scene in which sites communicate with different paces ranging from 1 to 32. Figure~\ref{fig:fed_E} represents that our proposed PointTransformerDDA+ is more robust compared to the base PointTransformer, which demonstrates the effectiveness of our proposed components. Even with a pace of 32, our model still yields an AUC of 0.738, while the base PointTransformer shows a large performance decrement.

\newpage
\section{Model Visualization}

\begin{figure}[!ht]
\centering
\includegraphics[width=0.45\textwidth]{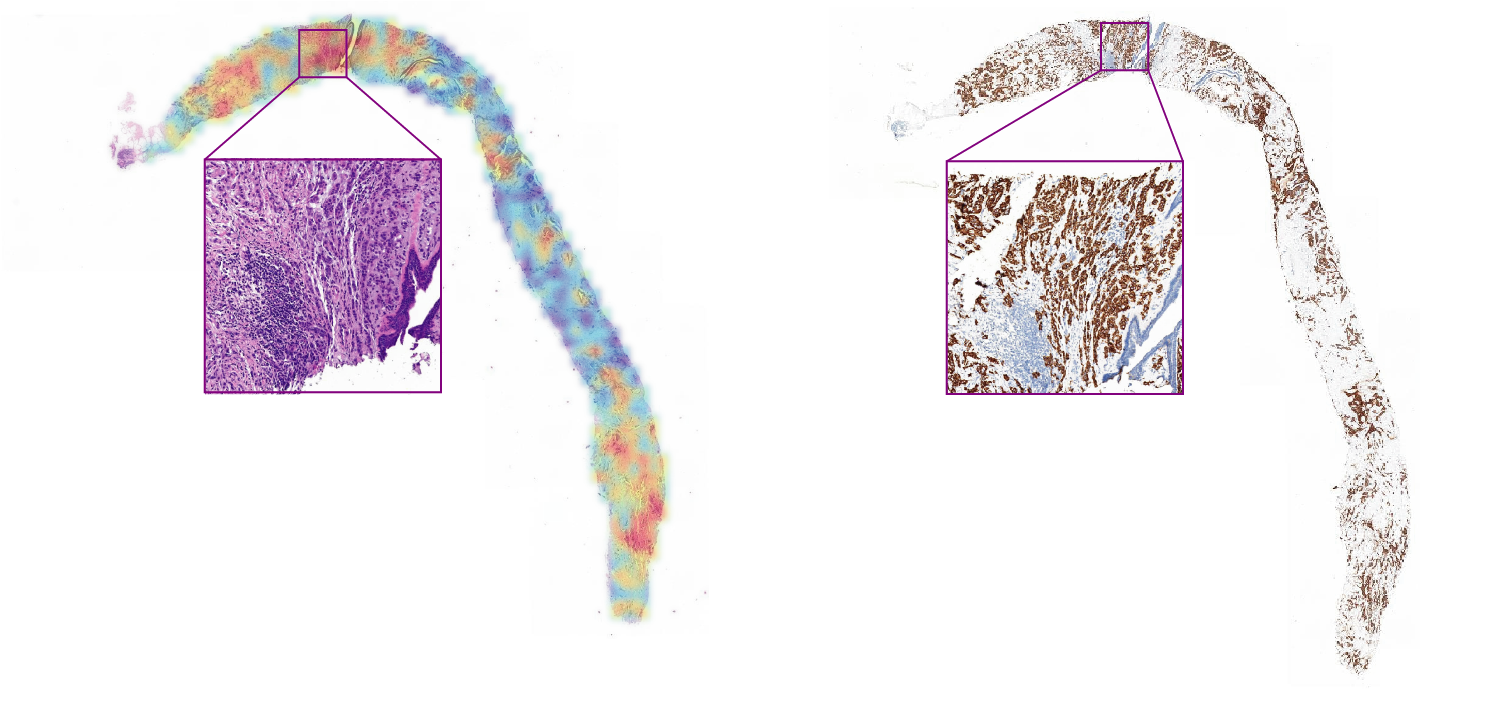} 
\caption{Visualization of a representative HE-stained WSI and its corresponding IHC-stained WSI with IHC score 3+.}
\label{fig:wsi_viz}
\end{figure}

We further try to visualize our model using integrated gradients~\cite{sundararajan2017axiomatic}. Generally pathologists should access the HER2 status from IHC-stained WSI as shown in the right part of Figure~\ref{fig:wsi_viz}. Deriving HER2 status from HE-stained WSI is challenging compared to IHC-stained WSI. However, our model can capture the deep IHC-stained area from HE-stained WSI as shown in the left part of Figure~\ref{fig:wsi_viz}.

\end{document}